\documentstyle[prl,aps,epsfig,preprint]{revtex}

\begin{document}                
\draft
\preprint{}

\title{Parametric Self-Oscillation via  Resonantly Enhanced 
Multiwave Mixing}
\author{A.~S.~Zibrov$^{1,4,}$\footnote{present address: National Institute 
of Standards and Technology, Boulder, Colorado, 80303}, M.~D.~Lukin$^{2}$, and
M.~O.~Scully$^{1,3}$}
\address{$^1$Dept.~of Physics, Texas A \& M University, College Station, 
TX 77843, $^2$ ITAMP, Harvard-Smithsonian Center for Astrophysics, Cambridge,
MA, 02138, $^3$Max-Planck Institut f\"ur Quantenoptik, 
D-85748 Garching, Germany, $^4$Lebedev Institute of Physics, Moscow, Russia}



\maketitle

\begin{abstract}
We demonstrate an efficient nonlinear process in which 
Stokes and anti-Stokes components are generated spontaneously 
in a Raman-like, near resonant media driven by 
low power counter-propagating fields. Oscillation of this kind does 
not require  optical cavity and  
can be viewed as a spontaneous formation of atomic coherence 
grating. 
\end{abstract}

Theoretical and experimental work of past few years on atomic coherence 
and interference has demonstrated a potential to improve significantly 
the existing nonlinear optical techniques \cite{nln}. In the present 
Letter we  demonstrate  efficient parametric generation 
accompanying the  {\it spontaneous} formation of the coherent 
superposition states.  Specifically, the present work reports the  
observation of the spontaneous parametric self-oscillation in resonant Raman 
media.  Such generation does not involve optical cavity and appears 
under remarkably simple circumstances, 
when two low-power counter-propagating fields interact with 
the medium. Oscillation manifests 
itself as  Stokes and anti-Stokes components generated with a frequency 
shift corresponding to that of the Raman transition. As the oscillator
goes over threshold, dramatic increase and narrowing of the beat note 
between the input field and generated components takes place. 

The principal possibility of mirrorless parametric oscillation with 
counter-propagating signal and idler fields has been
suggested in 1960's by Harris \cite{haold}. The original proposal based on 
non-degenerate frequency mixing has not been realized up to now due to
small values of nonlinearities in available materials and 
difficulties in achieving phase matching \cite{shen}. 
It is easier to achieve mirrorless oscillation in 
degenerate four-wave mixing. The possibility of self-oscillation 
in such interactions has been predicted in \cite{yariv}, and
a number of the related effects, such as conical 
emission or transverse pattern formation have been observed 
in a vapor driven by very strong, off-resonant counter-propagating laser 
beams \cite{many}. Workers in the field have also noted 
the importance of Raman nonlinearities in the early experiments 
on polarizations instabilities  \cite{boyd}.
 
As compared to the above work the presently
reported results utilize atomic coherence gratings \cite{phil,we} in a 
resonant double-$\Lambda$ atomic system (Fig.1a). Coupling of 
counter-propagating Stokes and anti-Stokes fields via such a
grating appears to be the main physical  mechanism resulting in 
Raman self-oscillation \cite{we}. Similar to several related 
studies \cite{nln,phil,hakuta,win} 
the present work operates in a so-called strong coupling regime in 
which nonlinearities can not be derived from
a usual perturbation expansion. In this regime quantum 
coherence and 
interference have a profound influence on nonlinear parametric amplification.
For example, linear and non-linear 
absorption of parametrically generated fields can be controlled and 
the phase mismatch, inherent in all 
non-degenerate parametric interactions involving counter-propagating
fields, can be  
easily compensated by a large dispersion accompanying
resonances in phase-coherent media. It is important that due to
quantum interference the strong-coupling regime was reached in 
the present  work with a very low driving power. The present results
are therefore directly related to recent theoretical studies on
few photon quantum control \cite{ima}, switching \cite{harris98}, and 
quantum noise correlation \cite{sq}, and can potentially be used to study 
interactions of a very low-energy fields and for suppression of quantum noise. 
Moreover, if extended to resonant molecular vapor the present approach 
might be useful for efficient Raman frequency shifting.  Likewise,
narrow-linewidth signals may also be of interest e.g. for optical magnetometry. 

The present experiments follows our 
previous work on atomic coherence effects in  optically driven $\Lambda$ 
systems in Rb \cite{win,ber}. 
Studying the detailed lineshape of these signals, we 
found that under certain conditions the Raman amplification \cite{ber}
can actually turn into 
a coherent self-oscillation of the Stokes and anti-Stokes components.
An essential element for the oscillation to appear is the existence of the 
two driving fields ($E_f,E_b$) propagating in the opposite directions.   
We first discovered this oscillation as 
the result of simple Fresnel reflection from the rear window of the Rb cell.   
Figure 1b shows the simplest experimental configuration that 
produced Raman oscillation.  A 
beam from an extended-cavity diode laser passes successively 
through an optical isolator (I), a focusing lens, a heated Rb cell, a
partially reflecting mirror and onto a fast photodiode (PD).  The signal from 
the photodiode is detected using a microwave spectrum analyzer (SA).  
The partially
 reflecting mirror (M) is used to retro-reflect some of the transmitted beam 
back through the Rb vapor.  With proper tuning of laser frequency, 
the backward beam 
causes the Raman gain peak to grow to the point of 
oscillation threshold.  When the self-oscillation occurs the 
detected Raman beatnote signal at  a frequency of hyperfine splitting  
($\omega_{hfs}$) increases in amplitude by as much as
  60dB and its linewidth narrows from  200 kHz to less than 300 Hz (Fig.2a).  
Under appropriate 
conditions the beatnote linewidths as narrow as 100 Hz FWHM were observed
(Fig.3a). This is  much
narrower than the usual broadening mechanisms for Raman transitions under 
the present conditions conditions (primarily, transit broadening $\gamma_{bc} 
\sim 50$ kHz and power broadening  $\sim$ 500 kHz).  The 
oscillation occurs without any cavity enclosing the cell.  We have been 
careful to eliminate
possible extraneous sources of feedback to lasers or other 
optical and electronic elements.  
     
In order to study the physical origin of the oscillation process 
we carried out 
a series of experiments, where instead of reflecting the incident laser 
light we injected laser beams with different frequencies from the opposite
directions (Fig.1c).  We found that 
the oscillation occurs readily if the forward and backward fields are 
tuned to the 
different ground state hyperfine level as diagramed in Fig.1a. It is 
more difficult to make the system oscillate if the backward 
beam is tuned to the same frequency as the forward beam. 
If tuned to different ground state hyperfine levels, the oscillation 
was observed with the backward beam coupling either the same ($P_{1/2}$) 
or different ($P_{3/2}$) upper-state fine-structure levels as the 
forward beam.  In our two-laser experiments it
is was easy to see oscillation for both $^{85}$Rb and $^{87}$Rb isotopes. When 
oscillating, the Rb vapor can convert as much as 4$\%$ of the total 
input power into the frequency shifted Stokes
and anti-Stokes sidebands. 

Typical conditions to observe the oscillation with a 
single laser 
beam are: ECDL  tuned in the wing of the Doppler profile of the $^{85}$Rb, 
D1 line (F=2 to F'=3) transition, $\sim$ 10 mW of power in spot size about 
$\sim 500$ microns, 5 cm long Rb cell operated at 75-95 C, and between 
10 and 80 $\%$
of the driving power is retro-reflected back through the Rb cell.  
In  the two-laser experiments oscillation 
was observed for driving input powers $2-10$ mW, spot sizes $0.1-2$mm, 
and cell temperatures $65-100 ^o$C. 

 The oscillation frequency shift ($\omega_0$)
does change somewhat with laser tuning 
(typical case was  30 Hz per MHz of laser tuning) and with angle between 
the forward and backward beams \cite{lowcomb}.  However, the oscillation 
frequency always 
remains within the bandwidth (few hundred kHz) of the power broadened and 
shifted single-beam Raman gain peak.  The oscillation prefers, but does
not necessarily require a circularly polarized beam, and the gain is 
largest with zero applied magnetic field. 

We have analyzed the characteristics of the 
forward and backward beams by making beatnotes with independently tuned 
laser sources, and by using optical cavities to analyze the spectra.  
We found that the field components at frequencies
of the forward and backward driving fields ($\nu_f$ or $\nu_b$) are 
surrounded by  generated first order Stokes and 
anti-Stokes fields at frequencies $\nu_{f,b}\pm\omega_0$. In certain 
cases second order components have been seen as well. The generated 
components produce, in general, asymmetric spectrum. In particular, 
in cases when forward driving field is tuned to e.g. upper ground state 
hyperfine sublevel and backward driving beam is tuned to the lower 
hyperfine sublevel, the anti-Stokes component observed in a forward direction 
is much more ($ \sim 20-30$dB) intense  than the Stokes one.  

These observations suggest that the actual oscillation mechanism is 
somewhat different from (although related to) that studied 
theoretically in Ref.\cite{we}.
That work involved only one pair of counter-propagating components. 
Motivated by experimental results, we consider a theoretical model in which
atoms in a double $\Lambda$-type configuration  
are interacting with six optical fields. These include
two counter-propagating driving fields  with frequencies $\nu_{F},\nu_B$ and complex slowly 
varying amplitudes 
${\cal E}_F$ and ${\cal E}_B$; anti-Stokes and 
Stokes components with frequencies 
$\nu_{1,3} = \nu_F \pm \omega_0$ propagating in the forward direction  
(${\cal E}_1,{\cal E}_3$), 
and corresponding components with frequencies $\nu_{2,4} = \nu_{B}
\pm \omega_0$ 
propagating in 
the backward direction (${\cal E}_2,{\cal E}_4$). The field is then written as
$E = \sum_{i} ({\cal E}_i {\rm e}^{-i(\nu_i t+ k_i r)} +{\rm c.c.})/2$.
Below we focus on the linear theory describing 
the oscillation threshold. Hence, all generated components are 
treated to first order only and saturation effects are disregarded.
These assumptions allow us to truncate the infinite 
hierarchy of equations. The resulting polarization can be written in the
form $P = \sum_{i} ({\cal P}_i {\rm e}^{-i(\nu_i t + k_i r)} + {\rm c.c.})/2$. 
We are interested here in polarizations at the Stokes and anti-Stokes 
frequencies, which are related to the field components by
$4\times 4$ susceptibility matrix $\chi_{mn}$:
${\bar {\cal P}}_{m} = \epsilon_0 \chi_{mn} {\rm exp}(i k_{mn} r) {\bar {\cal E}}_{n},$
where ${\bar {\cal P}}
= \left[\matrix{{\cal P}_1, & {\cal P}_2, & ({\cal P}_3)^*, & ({\cal P}_4)^*}\right]^T$,  
${\bar {\cal E}}
= \left[\matrix{{\cal E}_1, & {\cal E}_2, & ({\cal E}_3)^*, & 
({\cal E}_4)^*}\right]^T$, and $k_{ij}$ are representing free-space wave vector 
mismatch. For the present problem the matrix elements of 
$[\chi]$ were calculated explicitly for each velocity group and averaged 
over Maxwellian velocity distribution. In the present calculations we 
consider fields interacting in a slab of medium
of the length $L$. Assuming that the solution is homogeneous in transverse 
directions leads to $(k_{ij})_{\perp} = 0$,   
and the evolution along the longitudinal direction
$z$ is described by: ${\partial \over \partial z} {\cal E}_i = i/(2\epsilon_0) 
 (k_i)_z  {\cal P}_i$. The appropriate boundary conditions are taken to include 
a weak ``seed'' input
(${\cal E}$) at anti-Stokes frequencies (corresponding to e.g. spontaneous 
emission, or vacuum field). 

Before proceeding with comparison of experiment and theory 
we illustrate the origin of the oscillation. 
To this end, let us assume that absorption of the driving fields is 
negligible, and there is no inhomogeneous broadening. Furthermore, we 
disregard the coupling of the forward (backward) driving field with all 
transitions except for $c \rightarrow a$ ($b \rightarrow a'$) and  
assume that the detuning of the backward driving field  from 
respective single photon resonance ($\Delta_B$) is much larger that 
the corresponding detuning of the forward drive. In such 
a situation only forward anti-Stokes (${\cal E}_1$) and backward Stokes 
(${\cal E}_4$) fields are involved into nonlinear interaction (Fig.4a, \cite{we}). 
In this case: $\partial{\bar {\cal E}}_i/\partial z = a_{ij} {\bar {\cal E}}_j$, with 
$\{i,j\} = \{1,4\}$ and $a_{11} = -\eta_1 [\gamma_{bc} + 
i (\omega_{0}-\omega_{hfs})  + i 
(|\Omega_B|^2 -|\Omega_F|^2)/\Delta_B]/|\Omega_F|^2 - i k_{11}$, 
$a_{14} = i \eta_1 [ \Omega_B\Omega_F/(\Delta_B |\Omega_F|^2)]$, 
$a_{41} = i \eta_4 [ \Omega_B^*\Omega_F^*/( \Delta_B |\Omega_F|^2)]$,
$a_{44} = -i k_{41}$.  Here $\eta_i = (k_i)_z 3/(8 \pi^2) N (\lambda_i)^3 \gamma_i$,
where $\lambda_i$ is a wavelength of the $i{\rm th}$ field component and 
$\gamma_i$ is the radiative decay rate on the transition coupled by this 
component of the field. $N$ is atomic density, and $\Omega_{F,B}$ are 
Rabi-frequencies. 

When the phase matching condition is satisfied (${\rm Im} (\delta a) = 0$, 
$\delta a \equiv (a_{11} - a_{44})/2$), we find:
\begin{eqnarray}
{\cal E}_1(L) \sim {\cal E}_4(0)^*  \sim  {{\cal E} \over   \delta a  \; {\rm sin}(sL) - s \; 
{\rm cos}(sL)}, 
\label{solu}
\end{eqnarray}    
where $s = \sqrt{a_{14}a_{41} - (\delta a)^2}$, and the unimportant
proportionality constants have absolute values of the order of unity.
These solutions diverge if
$ {\rm tan}(s L) = s/(\delta a)$, 
which indicates the onset of mirrorless oscillations. Note that the latter
condition can be satisfied if  $\eta_4 
|\Omega_F\Omega_B| > \eta_1 \gamma_{bc} |\Delta_B|$, which is identical 
to a strong coupling condition of 
Refs.\cite{nln,phil,we,hakuta,win,ima,harris98,sq,ber}. Let us examine 
now the phase matching. 
Close to the two-photon resonance we have: 
\begin{eqnarray}
\kappa  (\omega_0 - \omega_{hfs} - \xi)  + 
c [k_f +k_b - k_1 - k_4]_z = 0,
\end{eqnarray}   
where $\xi=(|\Omega_F|^2-|\Omega_B|^2)/\Delta_B$ is a 
function of drive power and detuings representing a 
light shift, and $\kappa = c \eta_1 (k_1)_z/|\Omega_F|^2$.    
It is interesting that this equation resembles closely the 
frequency pulling equation of the usual 
laser theory, with frequency stabilization coefficient $\kappa$. 
The first term in the left-hand side corresponds to atomic 
dispersion, and the second describes 
the geometrical phase mismatch. This contribution is proportional 
to the Raman transition frequency (see inset to Fig.4) and also 
depends on relative angles 
between driving beams. Hence it  plays a role analogous to the 
cavity shift. Note, however, that under the typical oscillation conditions
stabilization coefficient $\kappa \sim c/(\gamma_{bc} L) \gg 1$ and 
the oscillation frequency is locked to the light-shifted Raman 
transitions frequency.

This implies that in the example considered above the physics behind the
oscillation  phenomenon is the coupling of the counter-propagating Stokes 
and anti-Stokes fields via spontaneously created atomic coherence $\rho_{bc}$.
 On one 
hand such coupling results in scattering of these fields into each other, 
thus forming an effective feedback. One the other hand, this process is also
accompanied by parametric amplification. When both of the effects are present  
self-sustained oscillation can occur.

Let us consider the influence of other nonlinear processes on 
mirrorless oscillation. When only a forward driving field is 
present nonlinear interaction results in the coupling between 
forward (or backward) propagating Stokes and 
anti-Stokes fields (Fig.4b) leading to  coherent Raman scattering
and amplification of the co-propagating  pair of fields in the vicinity 
of two-photon
resonance \cite{we,hakuta}. Oscillation is not possible in this case, since no
 effective feedback is present. However, when coherent Raman scattering exists
in addition to the coupling between counter-propagating Stokes and anti-Stokes
components, it can result in lowering the 
oscillation threshold.   
The process shown in Fig.4c represents a different 
type of parametric interaction. It leads to the scattering of the
counter-propagating  anti-Stokes (or Stokes) waves into each other, which 
does not change the total photon number of weak fields. Consequently, it 
alone can never lead to the oscillations. However, 
oscillations can emerge if 
in addition to the parametric energy exchange additional amplification 
mechanisms (e.g. coherent Raman scattering) are present.

In general, for the detailed comparison of the theory and 
experiment all of the six processes of the type shown in Fig.4 should be 
taken into account.  They give rise to simultaneous generation of all 
components in both directions.  To make a comparison we have solved the full 
system of propagation equations numerically, taking into account 
Doppler broadening, and propagation of all fields. The results 
(Fig.2b) show good qualitative agreement with 
experiments.

This work would not have been possible without active involvement and 
contributions of L.~Hollberg and V.~Velichansky. The authors warmly thank 
them as well as M.~Fleischhauer, P.~Hemmer, S.~Harris, A.~Matsko, 
V.~Sautenkov, and 
G.~Welch for useful discussions, and 
T.~Zibrova for valuable assistance.  We greatfully acknowledge the  
support from the Office of Naval Research, the National Science Foundation, 
and the Air Force Office of Scientific Research.




\begin{figure}[ht]
 \centerline{\epsfig{file=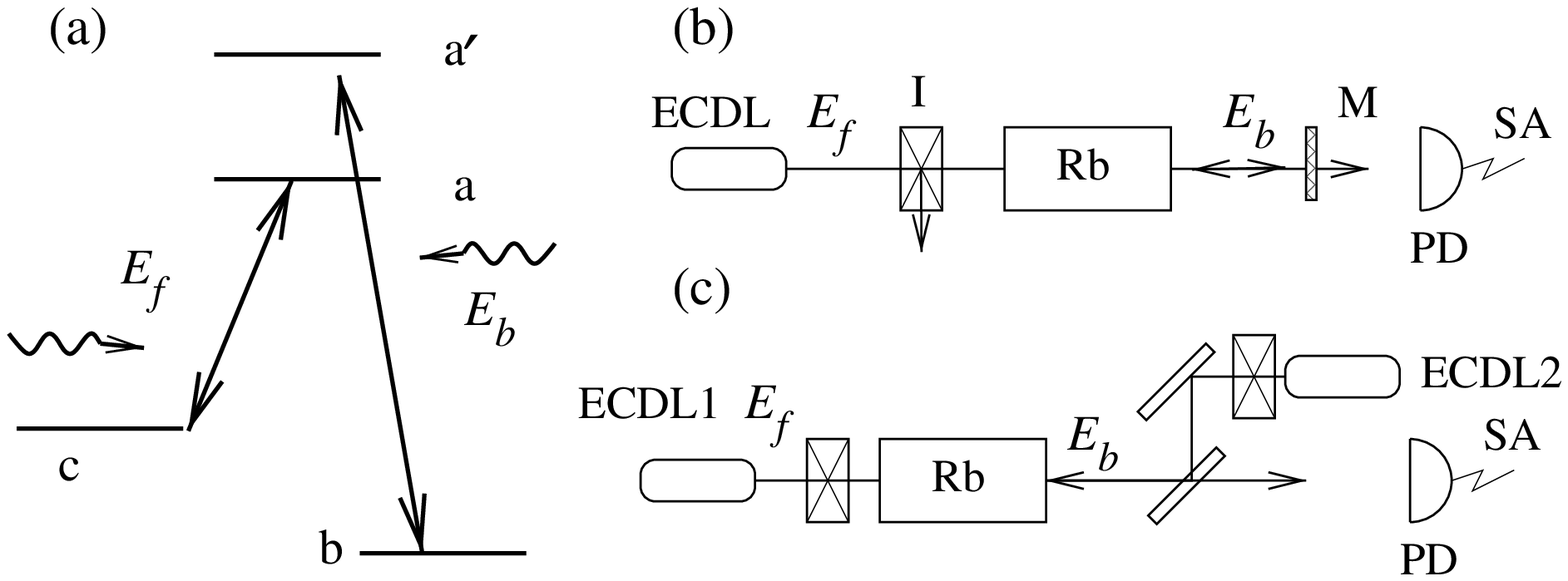,width=10cm}}
 \vspace*{2ex}
\caption{ (a) A prototype 4-level model for self-oscillations. 
In general, we assume that each driving fields couples both
of the ground states.  The upper levels of this double-$\Lambda$ 
system can represent some manifolds of states. In a particular case, $a$ and
$a'$ can also represent an identical state. In the experiment states
$c$ and $b$ are hyperfine sublevels of the Rb ground state $5S_{1/2}$.
(b,c) Experimental setups (schematic).}
\end{figure}


\begin{figure}[ht]
 \centerline{\epsfig{file=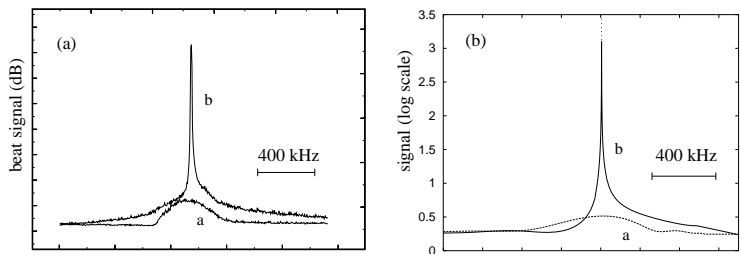,width=10cm}}
 \vspace*{2ex}
\caption{ a) A typical signal recorded by a fast photo-diode. Curve a is recorded with only forward driving beam present. Self-oscillations occur in the presence of the forward and backward driving fields (curve b). Parameters are: 
cell temperature 92 $^o$ C; forward driving beam with power 10 mW and 
spot size 
1.5 mm is detuned by 800 MHz to the red side of $F=3\rightarrow F'=3'$
transition of $D_1$ line; backward driving beam with power 2.5 mW and 
spot size 
1.5 mm is detuned by 2 GHz to the blue  side of $F=2\rightarrow F''=3''$
transition of $D_2$ line.   
b) Calculated signals
corresponding to the experimental conditions of Fig.2a. 
(At the point corresponding to parametric oscillation linear theory predicts 
infinite growth).}
\end{figure}


\begin{figure}[ht]
 \centerline{\epsfig{file=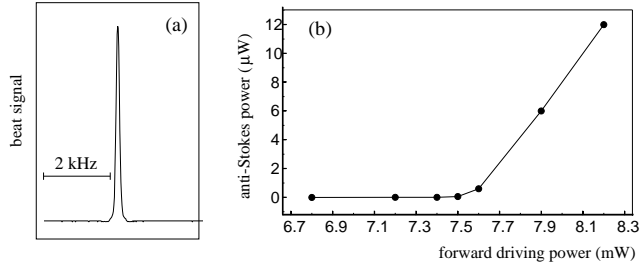,width=10cm}}
 \vspace*{2ex}
\caption{    
a) A typical beat 
signal at 3.034 GHz recorded when the frequency 
of the driving laser is locked to a reference cavity. b) A typical 
dependence of the generated anti-Stokes power as a function of the
input driving power in a vicinity of oscillation threshold.}
\end{figure}


\begin{figure}[ht]
 \centerline{\epsfig{file=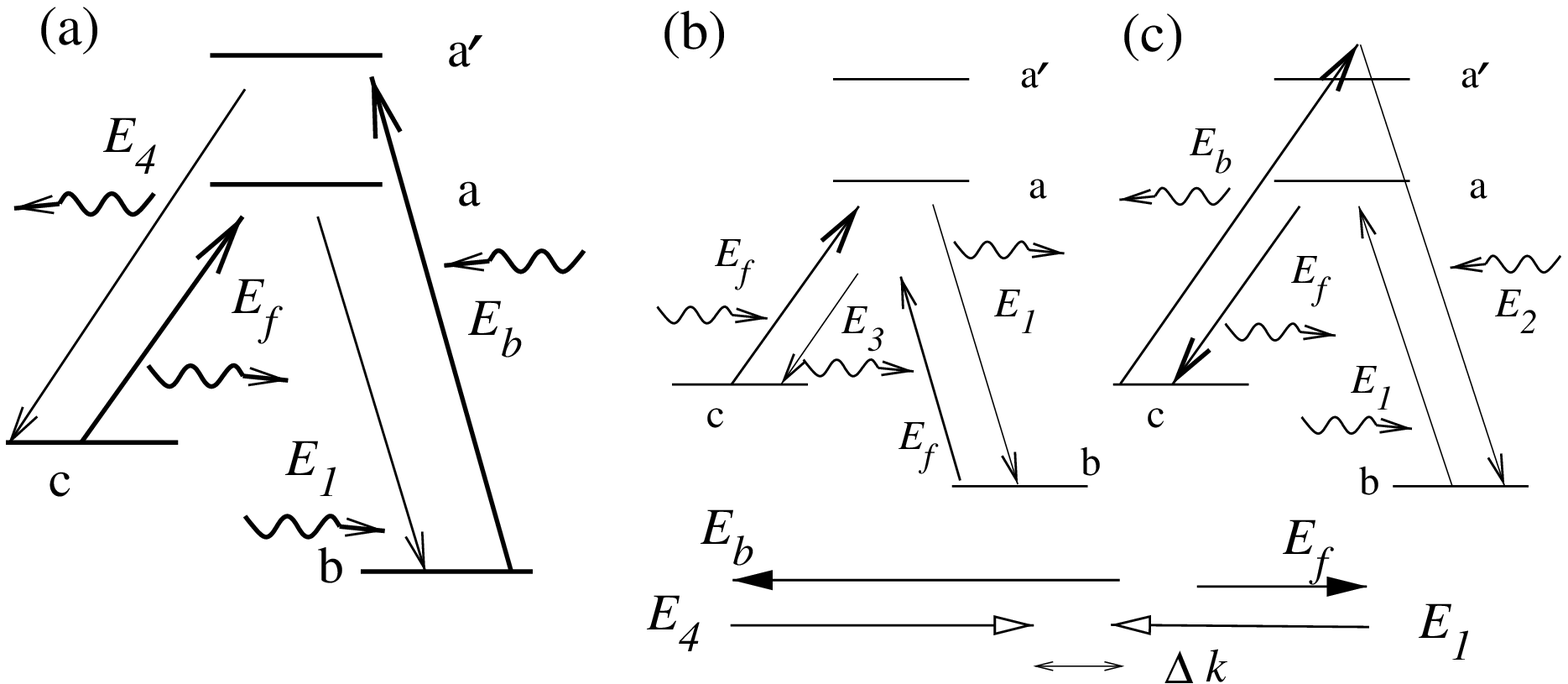,width=10cm}}
 \vspace*{2ex}
\caption{ Examples of different types of nonlinear processes contributing to 
self-oscillations. a) Direct nonlinear coupling between counter-propagating
Stokes and anti-Stokes fields; b) Coherent Raman scattering; c) Parametric
energy exchange between counter-propagating anti-Stokes fields. For each of 
the processes (a-c) there exists a complimentary process of the same type
involving other pair of weak fields (e.g. $E_{2,3}$ in Fig.4a). 
Inset illustrates wave vector mismatch 
for the process (a). If all fields are propagating
along the $z$ axes $[k_0^F +k_0^B - k_1 - k_4]_z = 2 \omega_0/c$.
}
\end{figure}



\frenchspacing

\begin{references}

\bibitem{nln} S.\ E.\ Harris, J.\ E.\ Field, and A.\ Imamo\u{g}lu, 
Phys.\ Rev.\ Lett.\ {\bf 64}, 1107 (1990); M.\ Jain {\em et al.},
Phys.\ Rev.\ Lett.\ {\bf 77}, 4326 (1996); for review see S. E. Harris, 
Physics Today {\bf 50}, \# 7, 36 (1997).  

\bibitem{haold} S.~E.~Harris, Appl.~Phys.~Lett., {\bf 9}, 114 (1968)

\bibitem{shen} 
Y. R. Shen, {\it The Principles of Nonlinear Optics} 
(John Wiley \& Sons, 1984).

\bibitem{yariv} A.~Yariv, D.~Pepper, Opt.Lett. {\bf 1}, 16 (1977).

\bibitem{many} See e.g. special issue on {\it Transverse Effects in 
Nonlinear-Optical Systems},  
J. Opt. Soc. Am. B {\bf 7}, 948-1157 (1990); {\bf 7}, 1264-1373 (1990);
M.Ducloy and D.Bloch, in {\it Optical Phase Conjugation}, p. 98, 
M.~Gower, and D.~Proch Eds., (Springer, 1994). 

\bibitem{boyd} D.~Gauthier {\it et al.},  
Phys. Rev. Lett. {\bf 61}, 1827 (1988); {\it ibid} {\bf 64}, 1721 (1990). 

\bibitem{phil} 
P.\ R.\ Hemmer {\em et al.}, 
Opt.\ Lett.\ {\bf 20}, 982 (1995). 


\bibitem{we} M.~D.~Lukin, P.~Hemmer, M.~Loeffler, and M.~O.~Scully, 
Phys. Rev. Lett. {\bf 81}, 2675 (1998).

\bibitem{hakuta} K. Hakuta, M. Suzuki, M. Katsuragawa, 
and J. Z Li, Phys.~Rev.~Lett. {\bf 79}, 209 (1997).

\bibitem{win} M.D.Lukin {\em et al.}
Phys.Rev.Lett. {\bf 79}, 2959 (1997).


\bibitem{ima} A.~Imamo\u glu, H. Schmidt, G. Woods, and
M. Deutsch, Phys.Rev.Lett. {\bf 79}, 1467 (1997); 
M. Dunstan {\it at al.}, in {\it Proc.: Quantum
Communication, Computing, and Measurement 2.} P. Kumar,
G.M. D'Ariano, and O. Hirota Eds.
  
\bibitem{harris98} S.~Harris and
Y.~Yamamoto, Phys.Rev.Lett. {\bf 81}, 3611 (1998). 

\bibitem{sq} M.D.Lukin {\em et al.} Phys.Rev.Lett. {\bf 82}, 1847 (1999).

\bibitem{ber} A.S.Zibrov {\it et al.}, in {\it Proc. 5th Simposium on 
Frequency Standards and Metrology'}, J.C.Bergquist Ed., 
(World Scientific, 1996).
 

\bibitem{lowcomb} When the gain is high it is possible to 
 have comb of oscillations with spacings between modes 
that could be tuned from 10's to 100's of kHz. This subsidiary comb 
structure depends strongly on angle of the return beam, the laser frequency, 
and is correlated with spatial mode changes in the generated light. 






\end{references}
\end{document}